Development and Performance of the Nanoworkbench:

A Four Tip STM for Electrical Conductivity Measurements Down to Sub-micrometer Scales

Olivier Guise[1,3,4], Hubertus Marbach[1,3,4,5], Moon-Chul Jung[1],

Jeremy Levy[2,4], Joachim Ahner[4,6], John T. Yates, Jr.[1-4],

[1] Department of Chemistry, [2] Department of Physics and Astronomy

[3] Surface Science Center

[4] Center for Oxide Semiconductor Materials for Quantum Computation

University of Pittsburgh, Pittsburgh, PA 15260

Tel: 412-624-8320, FAX: 412-624-6003

[5] present address: Universität Erlangen, 92058 Erlangen, Germany

[6] Seagate Technology, Pittsburgh, PA 15222

**Abstract**

A multiple-tip ultra-high vacuum (UHV) scanning tunneling microscope (MT-STM) with a scanning electron microscope (SEM) for imaging and molecular-beam epitaxy growth capabilities has been developed. This instrument (nanoworkbench) is used to perform four-point probe conductivity measurements at μm spatial dimension. The system is composed of four chambers, the multiple-tip STM/SEM chamber, a surface analysis and preparation chamber, a molecular-beam epitaxy chamber and a load-lock chamber for fast transfer of samples and probes. The four chambers are interconnected by a unique transfer system based on a sample box with integrated heating and temperature-measuring capabilities. We demonstrate the operation and the performance of the nanoworkbench with STM imaging on graphite and with four-point-probe conductivity measurements on a silicon-on-insulator (SOI) crystal. The creation of a local FET, whose dimension and localization are respectively determined by the spacing between the probes and their position on the SOI surface, is demonstrated.

# Introduction

In order to measure the electrical conductivity of surfaces and the influence of adsorbates on conductivity, it is desirable to perform local four-point conductivity measurements. The typical in-line set-up with four probes involves driving a current between the two outer probes while measuring the resulting voltage drop between the inner two probes using a high-impedance voltmeter (> $10G\Omega$ in our case) so that very little current is drawn. This yields a current-voltage (*I-V*) curve dependent on the separation *d* between the probes. It has a strong advantage over the two-probe measurements since it eliminates measurement error due to probe resistance and the contact resistance between the substrate and the probes. The four-terminal configuration also presents the advantage of being more versatile – provided that the probes can be controlled independently - allowing not only in-line measurements but also van der Pauw configuration electrical conductivity measurements in which the probes are placed in a square or rectangular configuration [1, 2].

Four-point-probe instruments are commercially available but the spacing between the probes is usually fixed and too large to be able to probe structures at micrometer or sub-micrometer levels. For this reason, it is desirable to implement instruments with smaller probe-to-probe distances.

Recently, the development of multiple-probe instruments for conductivity measurements has been reported [3]. Dual probe configuration measurements have been performed [4-7] as well as four-point probe measurements [8-10]. One reported instrument utilized four parallel insulating cantilevers – which cannot be driven independently - fabricated by a conventional silicon processing technique [11, 12]. Another approach is based on four independently-driven probes for conductivity measurements in ultra-high vacuum [13, 14].

We report the development and operation of the Nanoworkbench (NWB), a novel multiple-chamber ultra-high vacuum system based on a multiple-probe STM/SEM instrument designed to perform four-point-electrical measurements. Each of the four probes of the multiple-tip system can be operated independently as an STM with nanometer-scale lateral resolution and atomic-scale vertical resolution. This specificity greatly increases the potential of the instrument for conductivity measurement at sub-micrometer levels by allowing precise positioning of the probes and characterization of the surface or object of interest. The NWB also features a surface analysis and preparation chamber equipped with standard surface science tools as well as a molecular-beam epitaxy system.

## II. Overview of the System

A schematic picture of the basic structure of the Nanoworkbench (NWB) is shown in Figure 1a. Four STM nanomanipulators, oriented at 45° from the normal, and at 90° to each other, electrically probe a small region on the surface of a material of interest. They are guided by an SEM which images the surface and the tips. The sample is mounted on a positioning platform with X, Y and Z motion. Attached to the NWB are: (1) a chemical preparation and characterization chamber; (2) a molecular-beam epitaxy chamber (MBE); and (3) a fast-entry load lock chamber (see Figure 1b). Samples and probe may be transferred from these chambers to the NWB. The system was designed using AutoCAD 14 software (Autodesk, Inc.) in combination with ACIS 3D Open Viewer (Spatial Technology, Inc.). The entire system is mounted on an aluminum frame box designed and built as a highly rigid structure.

Each chamber is equipped with its own pumping system: a 500 L/s ion pump (Leybold), a 500 L/s magnetic turbo-molecular pump (Leybold, Pfeiffer and Shimadzu), and a titanium sublimation pump (Thermionics – model SB-1020). A common roughing line is used for the three experimental chambers. The pumps can be isolated from each other *via* a system of valves and gate valves. Pressure measurement in the system is achieved through cold-cathode gauges (MKS – model 421 cold cathode sensor), and pressure in the low $10^{-10}$ mbar range is routinely achieved for each chamber after bakeout. In addition, the load-lock chamber is pumped separately by a mechanical pump coupled with a 150L/s Leybold turbo-molecular pump. A pressure of $5.10^{-8}$ mbar is achieved within 1 ½ hour and an ultimate pressure of $6.10^{-9}$ mbar is readily achieved in the load-lock chamber without bakeout after 24 h.

## *II.1. Surface Analysis and Preparation Chamber*

As shown schematically in Figure 2 the surface analysis and preparation chamber has a spherical shape with 15" diameter, equipped with standard surface science tools, including a Perkin-Elmer Cylindrical Mirror Analyzer for Auger Electron Spectroscopy (AES model 10-155), a hemispherical Leybold-Heraeus EA-10 electron analyzer combined with a dual Mg/Al anode x-ray source for X-ray Photoelectron Spectroscopy (XPS), a UTI-100C quadrupole mass spectrometer (QMS), a home-made reverse view low-energy electron diffraction analyzer (LEED), an ion sputter gun and a gas dosing system. All the instruments are aligned radially along the chamber's equator (see Figure 2). An XYZ-manipulator with heating and cooling capability is used as a receiving stage for the sample box.

### II.2. Molecular Beam Epitaxy Chamber

The MBE chamber (12" in diameter) shown schematically in Figure 3 is used for thin film deposition, doping or growth of quantum dots. It contains three 3 kW solid-source electron-beam evaporators (MDC – evap3000) and one 100 W wire source (MDC – evap100). A Quartz Crystal Microbalance (QCM – Sycon model STM-100) is used to monitor thin film growth. An XYZ-manipulator similar to the one used in the preparation chamber serves as the sample box receiver. Each evaporation source can be isolated *via* a shutter mounted on a linear feedthrough.

### II.3. STM/SEM Chamber

The STM/SEM chamber is equipped with four nanomanipulators (Kleindiek Nanotechnik—model MM3 and MM3a) holding etched PtIr probes. The addition of a UHV-SEM (FEI—model 2LE-EVA) based on a Schottky emitter source allows fast and precise approach and positioning of the probes. The STM/SEM chamber also features a UTI-100C quadrupole mass spectrometer, a gas handling system as well as a probe-garage holding up to 10 tips and a sample box carousel (3 sample spots).

## III. Transfer System and Sample Box

### III. 1. Sample Box and Receiving Stages

For the preparation of clean Si(100)-2x1 surfaces [15, 16], it is important to have precise control over the temperature of the sample in each UHV chamber. Moreover, it is desirable to have a system which can accommodate samples with various sizes and requirements. For this purpose a special transferable sample box with integrated heating and temperature measuring capabilities has been developed. As shown in Figure 4 the sample box is made of OFHC copper with highly-polished outer walls. Four UHV-compatible male banana plugs (Transfer Engineering and Manufacturing - model BPBPM) are tightly connected to the sample box and isolated from it with ceramic spacers. The outer plugs are made one of chromel and the other one of alumel. They are used for K-type thermocouple connections. The inner plugs are used as electrical heating connections and are made of BeCu, 304 stainless steel, and gold-coated for enhanced electrical performance and reliability. The sample box is equipped with stainless steel inserts on each side. These inserts with tapered holes match tapered pins on the transfer manipulator and are used to grab the sample box for UHV-transfer.

The sample box can be equipped with either resistive or radiative heating devices. Figure 4 shows the resistive heating setup with a 10 mm x 5 mm x 0.5 mm Si(100) crystal. The crystal is resistively heated using non-interacting Ta contacts following a method described by Yates [17]. Alternatively, the sample box can be equipped with a radiative heater (Tectra – Boralectric – not shown here).

Each UHV chamber is equipped with a receiving stage such as the one shown in Figure 5. The receiving stage is also made of OFHC copper - with highly-polished inner walls closely matching the outer walls of the sample box - and is equipped with a similar set of female banana plugs. The receiving stage features a set of 2 mm diameter stainless steel pins on each side to ensure rigorous alignment of the sample box during transfer. In addition, each receiving stage features a liquid nitrogen reservoir for cooling the sample.

## III. 2. Transfer System

The transfer system is articulated around two main axes: Axis 1: Load-lock chamber – Preparation chamber – STM/SEM chamber, and Axis 2: MBE chamber - Preparation chamber, as can be seen in Figure 1a. Linear/rotary manipulators are used to transfer the sample throughout the system. In addition, two wobble-sticks are mounted on the STM/SEM chamber to allow for easy transfer of the sample box to the STM stage sample carousel as well as *in situ* tip-exchange. We use custom-made magnetically-actuated wobble-stick manipulators (Ferrovac) which are not affected by the pressure difference between atmospheric pressure and ultra high vacuum, allowing the shafts of the wobble-stick to be guided in a smooth and controlled way. Both linear/rotary manipulators and wobble-sticks are equipped with a stainless steel fork system (pincer/grabber) featuring jaws with two stainless steel tapered pins at the end. These pins match the stainless steel inserts with tapered holes (see Figure 4) in the sample box and ensure enhanced stability during transfer of the sample box.

The typical transfer sequence is divided into 4 steps (see Figure 5):

1) The linear transfer manipulator holding the sample box approaches the XYZ manipulator.
2) Once the sample box is precisely positioned above the receiving stage, the XYZ manipulator is translated up, so that the receiving stage houses the sample box.
3) Once the connection between the sample box and the receiving stage is tight, the fork system (pincer) at the end of the transfer manipulator is opened and,
4) The linear manipulator is moved away from the sample box.

# IV. Multiple-tip STM/SEM Chamber

## *IV. 1. Overview - Nanomanipulators and XYZ-table*

The STM/SEM chamber is directly connected to the preparation chamber (see Figure 1a). Its core is a multiple-tip scanning tunneling microscope (MT-STM) combined with a scanning electron microscope (SEM).

The MT-STM assembly is shown in Figure 6 (6a: side view and 6b: top view). It comprises four nanomanipulators models MM3 and MM3a (Kleindiek Nanotechnik) mounted on a stainless steel ring at 90° from each other in the horizontal plane (see Figure 6a) and inclined at 45° from the sample plane (see Figure 6b). This design ensures a large opening from the top to accommodate the SEM and its detector. Electronic components such as the preamplifiers for STM feedback control and imaging are also mounted on the STM stage mounting ring (Figure 6a) and are connected to the outside *via* electrical feedthroughs.

Both the STM nanomanipulators and the sample stage must be capable of coarse and fine motion. The design of the Nanomotor ® [18] achieves both through its unique design. For fast large travel, the Nanomotor ® can be driven in the coarse step mode which exploits the well known stick-slip mechanism [19]. At the same time, positioning in the nanometer range is possible in the fine step mode, which utilizes the piezoelectric effect in the conventional way [20].

The sample table (Kleindiek Nanotechnik, model LT6820XYZ) has three perpendicular degrees of freedom and is equipped with three Nanomotors. The table can travel 20 mm in the X and Y directions and 15 mm in the Z direction. The integrated optical encoding system allows positioning of the sample with 100 nm repeatability. The stiffness of the table was enhanced by mounting additional vertical aluminum columns.

The nanomanipulators have been developed in collaboration with Kleindiek Nanotechnik for STM applications. As shown in Figure 7 the nanomanipulators have two rotational axes (X, Y) and one linear motion (Z). Two slightly different models are being used: the MM3 model with a length of 80 mm and the MM3a model with an overall length of 65 mm.

From the standpoint of control electronics, the simultaneous operation of four STM devices or the performance of four point probe conductivity measurements on a small scale is a challenge by itself. The SPM electronics for the 4-tip STM is provided by Kleindiek Nanotechnik (Nanomotor® STM Control and Image Acquisition Electronics / Software). The electronics is controlled *via* a computer through an RS-232 serial interface. Each SPM electronics system is controlled by a dedicated PC to guarantee simultaneous independent operation of all four nanomanipulators. An analog feedback-loop is used in constant-current mode. To reduce coupling of ambient electromagnetic noise the first amplification stage of the tunneling current is located inside the UHV system close to the nanomanipulators using a low-noise fast operational amplifier OPA111 (Burr-Brown). The gain of the OPA111 is set with a 100 MOhm resistor. With this setup the range of voltage/current is -2.5 V to +2.5 V (restricted by SPM electronics)/ -100 nA to +100 nA (restricted by the specification of OPA111 in combination with a 100 MOhm resistor). To enable four-point conductivity measurements, direct electrical access to the probes is also desirable. Low dissipation electromagnetic relays (Teledyne, type 732-12) are mounted close to the OPA111, as shown in Figure 8.

As shown in Figure 7, the MM3 nanomanipulator is composed of a tube connected to the piezo drive, whose inner diameter is 1.50 mm. The tip holder must fit in this tube and hold onto it. The tip holder is made out of stainless steel. Tip exchange under UHV requires the use of wobble stick to grab, manipulate and transfer the tip holder assembly. As shown in Figure 9, the tip holder is slotted parallel to its central axis (7mm deep) so that its upper part acts as a spring. When grabbing the tip holder with a wobble-stick, the upper part is compressed, reducing the effective holder diameter, allowing the tip to be inserted into the manipulator tube. Upon releasing the tip holder, the force of the spring is released, holding the tip assembly in position.

## *IV. 2. UHV-SEM + MCP Detection*

The SEM column (FEI - Model 2LE EVA) is based on a Schottky emitter source with a two-lens electron technology and an electronically-variable aperture. Mounted on top of the chamber, it is mainly used to locate features on the surface and to guide the STM tips near objects of interest.

Conventional scanning electron microscopes (SEM) operate at pressures near $10^{-6}$ mbar. Therefore typical SEM detection systems are based on scintillation amplifiers (a summary of SEM detectors can be found in [21]). In these systems secondary electrons are converted into light and subsequently the photons are converted back into an electrical signal, which limits the sensitivity and signal-to-noise ratio (typical gains are in the range of $10^4$).

The UHV environment of the nanoworkbench system permits the use of a high efficiency multi-channel plate (MCP) system, similar to those used in UHV instruments for ion and electron spectroscopy [22]. A typical MCP consists of an annular array of tiny glass tubes (12-23 micrometer diameter and 0.5 to 1mm long) fused together to form an array of thousands of independent electron multipliers. An accelerating potential is applied to a metal film deposited across the ends of the tubes, and when an electron enters a tube it produces additional secondary electrons as a result of striking a special highly emissive coating on the inside of the tube. Two MCPs stacked together are used to create a gain approaching $10^7$.

Figure 10 shows a schematic diagram of the ultra high efficient secondary electron detection, the electron beam deflection and image recording system used in the nanoworkbench.

Here we use an ultra-thin (2.5 mm maximum) annular MCP assembly with a center hole (Hamamatsu F6589), which is directly mounted below the SEM electron column. The shielded center hole allows undisturbed travel of the primary electron beam to the sample surface. The detector position above the specimen surface is ideal for collecting a large percentage of secondary and backscattered electrons emitted from the specimen surface. Because of the large solid angle detection efficiency of the MCP, shadowing due to the nanomanipulators and the topography of the specimen is minimized; even deep holes and vias can be seen with greater efficiency than with side mounted secondary electron detectors. In particular, in the area of critical dimensions, the MCP can provide uniformly illuminated shadow-free images from which highly accurate measurements can be obtained. The front surface of the detector array may be biased positively or negatively to attract or repel secondary electrons from the sample. In this way a secondary electron image or a backscattered electron image may be selected easily.

The electronic amplification and pulse shaping of the MCP output signal is performed by a combination of a commercial available preamplifier (ORTEC Fast Preamplifier model 9301), and an amplifier (ORTEC Amplifier Discriminator model 9302) along with an home-built pulse shaper. The resulting pulse sequences (up to 20 MHz frequency) are digitized and synchronized with the deflection controller using a commercially available, fully programmable scan system (ADDA II, Soft Imaging System). The ADDA II system is connected to a PC for control of scan parameters and image recording.

Under experimental conditions – in four-point probe configuration - the SEM resolution is approximately 100 nm. Large working distances imposed by the design of the STM stage limit the SEM-resolution of the instrument.

### *IV. 3. Docking-stage*

The STM unit is decoupled from the electron column mounted on top of the STM/SEM chamber *via* the spring isolation system and the eddy-current damping unit (see section IV. 4). As a result, the distance between the sample (which floats during STM imaging) and the electron gun aperture is not constant. Such variation in the working distance makes it difficult to focus the SEM, especially at high magnifications ($\geq 10^3$).

A docking system has been developed to address this problem. Figure 11 shows the design of the docking mechanism. It consists of three pairs of metal pillars and a piston-like linear actuator. One metal pillar is mounted on the STM unit, whereas the second pillar in a given pair is mounted on the frame under the STM unit. The mating faces are mirror-finished to ensure excellent mechanical contact. The actuator is composed of a piston head and a cylindrical chamber without any contact with one another. These two elements are connected to the linear motion feedthrough and the STM unit, respectively. When the piston head of the actuator is pulled down (using the linear motion feedthrough) it makes contact with the cylindrical chamber to transmit the linear motion and, consequently, pulls down the STM unit until both pillars are in contact. A spring between the linear motion feedthrough and the actuator regulates the linear force to give a consistent pulling force. When the STM unit is undocked, the piston head does not have any contact with the cylindrical chamber, and the STM unit is floating, completely decoupled from the rest of the system (Figure 11a). This position is used for STM operation and conductivity measurements. For high-resolution SEM imaging, it is preferable to dock the stage (Figure 11b).

## *IV. 4. Vibration Isolation*

Vibration isolation is especially challenging since the NWB is a large system with UHV-requirements. Nonetheless, vibration isolation is crucial not only for STM operation but also for conductivity measurements since one would like to achieve stable contact of the probes with the substrate over a long period of time.

As a first step towards vibration isolation, the main mechanical pump is isolated by coupling the rough vacuum line to a heavy concrete block. In addition, each turbo-molecular pump is connected to the rough vacuum line via flexible tubing to minimize vibration transfer. During STM operation and conductivity measurements, all turbo-molecular pumps are turned off. Additional isolation of the entire NWB from the building is achieved through a six-point commercial active vibration system (HWLbioanalytic SYSTEMS, MOD-4). Each of the six elements senses vibrations along both horizontal and vertical directions, and a closed-loop feedback system compensates these disturbances. The isolation elements are active within the range of 0.75 – 35 Hz and passive above that. Additional vibration isolation elements are mounted inside the STM/SEM chamber.

### IV. 4. a. Spring Isolation and Eddy-current Damping

The STM assembly is suspended from the chamber with custom UHV-compatible springs, fabricated using Inconel X750 wire, 0.041" diameter (Gibbs Wire and Steel Co., Inc.). The whole STM assembly (four nanomanipulators, sample box receiver and XYZ-table) is suspended on four springs, each adjusted to have similar force constants (see Figure 6a). The springs were electropolished to reduce their porosity and limit their outgassing in vacuum by immersion in a 3:1 mixture bath of phosphoric acid and sulfuric acid [23], and a 9V potential with respect to an array of graphite electrodes was applied for 3 minutes. Every 60 seconds, the side of the springs facing the cathode was rotated by 120° to produce a uniform surface finish. Because the surface area of a spring is large, the ohmic drop across the length of a spring often produced an insufficient surface finish at the end. A conducting backbone was therefore used to keep the spring at the same potential throughout its length. Stock removal from the process is almost negligible, the diameter of the spring was found to decrease by a mere 0.001" upon electropolishing. The measured spring constant is 110.3 kg.s$^{-2}$ for each spring used.

To further dampen vibrations in the STM stage, an eddy-current system was employed, consisting of a circular array of samarium-cobalt magnets (Magnetic Component Engineering, Inc., model S2669, 1T) with interposed conducting copper blades. The copper blades are mounted on the STM unit and the magnets are mechanically clamped on the frame (Figure 6a). The samarium cobalt magnets have good corrosion resistance and high temperature stability. The Curie temperature is about 800 °C and the maximum working temperature is 320 °C, well above the requirement of our system for bakeout.

**IV. 4. b. Vibration Analysis**

Vibration analysis on the spring and eddy-current damping system were performed to ensure maximum effectiveness. The vibrational motion was created by simply lowering the STM stage and releasing it. The motion was recorded with a CCD camera interfaced to a computer. This analysis enabled us to optimize the force constant of the spring system as well as the damping constant. The resonance frequency of the vibration isolation system (~1.5Hz) could be kept far from the lowest natural frequency of the nanomanipulators (~300Hz), leading to an effective vibration isolation. Figure 12a shows the vibrational motion of the spring system without the eddy-current damper. The resonance frequency, $f_{meas}$=1.9Hz, is in good agreement with the calculated resonance frequency, $f_{calc}$=2.1Hz. Figure 12b shows a plot of the motion with the eddy-current damper. The damping factor, $\varsigma$, goes from 0.0419 without the eddy-current to 0.184 with the eddy-current system.

# V. Performance

The Nanoworkbench is capable of performing measurements in a variety of operational modes, some of which are illustrated below.

## V. 1. STM

The possibility of performing STM measurements distinguishes the nanoworkbench from other four probe setups. STM operation also expands the imaging capabilities of the NWB.

### V. 1. a. Sample Preparation

A highly-oriented pyrolytic graphite (HOPG) sample of 10mm x 10mm x 2mm (SPI – model 440HP-AB) is used for testing the STM capability of the nanomanipulators. The usual approach to prepare the HOPG sample for STM is to take a piece of tape, press it onto the flat surface and pull it off. The tape takes with it a thin layer of HOPG, leaving a freshly cleaved HOPG surface.

### V. 1. b. STM-operation and STM Imaging

The specifications of the Nanoworkbench STM probes differ in several important respects from typical single-probe STM instruments. First, as mentioned earlier, the manipulators are mounted at 45° from the surface plane. Therefore the Z motion of the manipulators is established with a similar angle and is not perpendicular to the sample surface as is customary for STM systems. The SPM software incorporates a tilt correction feature to compensate for the large angle. Improved STM operation was achieved by bending the tips ~25 degrees downwards (in direction of sample) at a distance of 1 mm before the end of the tip. Secondly, the nanomanipulators themselves are less compact and rigid than most STMs. Therefore the lowest resonance frequency of the nanomanipulators is close to 300 Hz. Despite the described procedures for vibration isolation and the increase of the overall stiffness built into the STM stage, slow scanning (~1Hz) was required to acquire satisfactory STM images.

Using custom-made PtIr etched probes (Custom Probes Unlimited) we were able to reproducibly image atomic steps (see Figure 13) and small features with a lateral width as small as 4 nm on a cleaved HOPG surface. Similar images were acquired with all four nanomanipulators.

## V. 2. Four-Point Probe

As explained above four terminal measurements have several advantages over two-terminal measurements. The nanoworkbench allows *in situ* positioning of four nanoscale contacts, allowing four probe measurements to be obtained.

**V. 2. a. Set-up and Strategy**

To perform the measurements the precise placement of the tips is crucial. The following procedure is employed for positioning the four probes:

1. The tips are brought into close proximity to the sample surface, using the UHV-SEM as a guide. The "electron shadow" of the probes in the SEM images can be used to estimate the distance between the tip and the sample.

2. When the distance between the tip and the sample is small enough the automatic STM Z coarse-approach with the feedback loop control is activated.

3. Steps 1 and 2 are repeated until all four tips are in tunneling contact with the surface. The spacing between the probes can be measured to an accuracy of ± 100 nm from the SEM image. Fine positioning of the tips in tunneling contact is possible.

4. Gentle contact is achieved between the tips and the sample surface. By changing the tunneling parameters (increase of current setpoint, decrease of bias voltage) the tunneling gap is minimized. By setting the gain of the feedback loop (SPM software) to zero, the feedback control is switched off. To prevent damage and/or interference of the operational amplifier (OPA111) all connections to the preamplifier are disabled via a switchbox. By switching an electromagnetic relay (see Figure 8), an electrical shortcut to the tip is established and the resistance between tip and sample can be monitored. With all four tips in electrical contact with the sample, four-probe measurements can be performed. A Keithley 6487 picoammeter/voltage source is used as a current source between the outer probes while the voltage drop between the inner probes is measured with an Agilent 34401A 6 ½ digit multimeter (input impedance >10 GOhm).

**V. 2. b. Sample preparation**

Four probe experiments were performed on silicon-on-insulator (SOI) samples (Ultrasil, Inc.), which consist of an undoped Si(100) layer with a thickness of 3.5 μm, separated from a highly doped Si(100) substrate called a "handle" (410 μm-thick) by a buried oxide layer (500nm-thick). The handle is highly doped (resistivity of 1-10 Ω·cm) allowing the insulating top layer to be flash cleaned by resistive heating of the handle (see schematic of the SOI sample in Figure 14b). The dimensions of the crystal are 10x5x0.5 mm$^3$. Prior to introduction in vacuum the crystal was chemically-cleaned *ex-situ* using the following recipe: (1). 10 minutes acetone; (2). 10 minutes ($H_2O_2$:$H_2SO_4$ 1:2) at 130°C [24]; (3). Standard Clean-1 at 65°C for 5 minutes ($H_2O$:$H_2O_2$:$NH_4OH$ 5:1:1); (4). Standard Clean-2 at 65°C for 10 minutes ($H_2O$:$H_2O_2$:HCl 6:1:1)[25, 26]; and (5). Oxide etch for 15 seconds using (HF:$H_2O$ 2:100). This treatment is known to produce a hydrogen-terminated Si(100) surface free of organic contaminants[27]. The crystal was subsequently annealed in vacuum for 20 minutes at 870 K, followed by 20 minutes at 1170 K and then by a 15-second flash at 1470 K. Following this treatment the Auger spectrum reveals an intense Si peak at 92 eV and only tiny signals from carbon (272 eV) and oxygen (502 eV) (not shown). The C/Si peak-to-peak ratio of the clean sample was found to be less than 0.002, which corresponds to a C atomic fraction of 0.005 in the depth of Auger sampling.

### V. 2. c. Results

A typical experimental configuration is shown on Figure 14b. In a first step, all four tips were brought in contact with the clean Si(100) crystal, according to the positioning procedure described above. The four probes are arranged in a linear fashion as shown in the SEM image Figure 14a. The spacing $d_1$ between the inner probes and $d_2$ between the outer probes was determined by SEM. Throughout the experiment $d_1$ and $d_2$ were maintained constant respectively at 2.0µm and 10 µm. A voltage ($V_{applied}$) was applied between the outer tips #1 and tip #4, and the resulting voltage drop between tip #2 and tip #3 ($V_{measured}$) was measured for different values of the handle voltage $V_{handle}$.

Figure 15a and Figure 15b summarize the resulting charactistics of the "device" formed by the SOI sample, the handle and the four probes. Figure 15a shows an intensity map of $V_{measured}$ as a function of $V_{handle}$ and $V_{applied}$. The characteristics are linear for low values of $V_{handle}$, and become highly nonlinear at sufficiently large $V_{handle}$. Figure 15b shows a "cross-section" of Figure 15a for 7 values of $V_{handle}$: -8V, -5V, -4V, 0V, +4V, +5V and +8V. The handle voltage causes a local depletion region at the "source" and "drain" which are the outer tips #1 and #4, thus reducing the measured value of the voltage. This transition from Ohmic behavior to a highly non-linear device characteristic is typical of a field effect transistor (FET). The difference is that this FET is "roaming" because the source and drain leads are STM probes that can touch down at different locations on the sample. A roaming FET may prove useful in probing the transfer characteristics of a variety of nanoscale devices such as quantum dot-based single electron transistors.

## VI. Acknowledgments

This work was supported by DARPA QuIST through ARO contract number DAAD-19-01-1-0650, a DURIP grant from ARO and a DARPA grant (DAAD-16-99-C-1036). Additionally, we express our gratitude to Thomas Gasmire from the machine-shop of the Department of Chemistry at the University of Pittsburgh for the construction and maintenance of the UHV-system and to Robert Muha of the electronics shop for design and maintenance support. We would also like to acknowledge the collaboration of Kleindiek Nanotechnik for the development of the SPM software and hardware.

**Figure Captions**

**Figure 1**: (a) Schematic picture of the Nanoworkbench. (b) Overview of the Nanoworkbench showing a multiple-tip STM / SEM chamber (left), a surface analysis and preparation chamber in the center featuring standard surface science tools and a molecular-beam epitaxy chamber (MBE) on the right equipped with Ge, Si and C evaporation sources. Additionally, a load-lock chamber allows for fast and easy transfer of the samples and probes following two transfer axes – Axis 1: Load-lock / preparation chamber / STM-SEM chamber and Axis 2: MBE chamber / preparation chamber.

**Figure 2:** Configuration of the surface analysis and preparation chamber. The chamber is equipped with standard surface science tools such as AES, LEED, XPS and QMS as well as a gas-handling system.

**Figure 3**: Top view of the evaporation system in the MBE chamber. Three evaporation sources (silicon, germanium and carbon) employ three separate shutters to control the evaporation process. A barrier separates the sources to prevent cross-contamination.

**Figure 4:** (a) Sample box equipped with four UHV-compatible banana plugs (models BPBPM) for heating and temperature measurements. Stainless steel inserts are used to tightly grab the sample box for transfer from one chamber to the next one. A resistive heating device is mounted on this sample box.

**Figure 5:** Snapshot of the sample box and receiver box in the preparation chamber. The transfer sequence is divided into 4 steps. (1) The sample box – held by the fork system of the linear/rotary manipulator approaches the XYZ receiver; (2) Once the sample box is positioned right above the receiving box, the receiver (controlled by an XYZ manipulator) is moved up, so that the receiving box tightly houses the sample box; (3) Once the connection is tight, the fork system is opened and; (4) the linear manipulator is moved away from the sample box. The sample box and receiving box are both made of highly-polished OFHC copper.

**Figure 6:** (a) Overview of the STM stage in the STM / SEM chamber. The STM stage features four nanomanipulators that can be used as STMs (models MM3 and MM3a – Kleindiek Nanotechnik) mounted on a ring at 45° from the sample plane, a sample stage mounted on an XYZ-table, as well as spring isolation and an eddy-current damping system. (b) Top view (SEM point of view) showing the four nanomanipulators mounted at 90° from each other in the sample plane (horizontal plane). The large solid angle allows enough space to accommodate the SEM-gun.

**Figure 7**: Left: a photograph of the MM3a nanomanipulator. Right: schematic showing both rotational axes (X, Y) and the linear motion (Z) of the nanomanipulator.

**Figure 8**: Schematic of preamplifier and relay electronics close to tip. The parts shown below the dashed line are located inside the UHV chamber.

**Figure 9**: Schematic of the tip holder assembly. The tip holder slides into the piezo tube of the nanomanipulator. PtIr etched wire is mounted in the tip holder.

**Figure 10**: Schematic of the SEM-MCP detection system.

**Figure 11:** Schematic view of the docking system. The STM unit is connected to the docking system without any mechanical contacts to the frame, (left) undocked and (right) docked. The undocked position is used for STM and conductivity measurements applications while the docked position is used for SEM imaging to keep the working distance constant.

**Figure 12:** Vibration analysis showing the relaxation of the STM unit after a manual perturbation, (a) without the eddy-current damper (resonance frequency = 1.9Hz, damping factor ς=0.0419) and (b) with the Eddy-current damper (resonance frequency = 1.9Hz, damping factor ς=0.184).

**Figure 13**: Reproducible STM imaging of atomic steps on HOPG acquired with MM3a nano-manipulator. The images were acquired consecutively. Experimental conditions for both images: 1x $10^{-8}$ mbar; $U_{bias}$ = 200 mV, $I_{setpoint}$ = 0.43 nA, image size 271x 271 nm, acquisition time per image was 260 s. The bar scale represents 50 nm.

**Figure 14**: (a) SEM image showing the in-line configuration with the inner tips at a distance of 2.5 $\mu$m. The insert in the upper right corner is a magnified view of the marked area. The scale bar represents 200 $\mu$m for the overview image and 10 $\mu$m for the insert. (b) Schematic of the experimental setup.

**Figure 15**: Four-point probe measurement on SOI sample. (a) Map of the voltage drop measured between the inner probes ($V_{measured}$) as a function of the voltage applied between the outer probes ($V_{applied}$) and the bias voltage on the handle ($V_{handle}$) and (b) Cross section of (a) for various values of $V_{handle}$.

# Overview of the Nanoworkbench

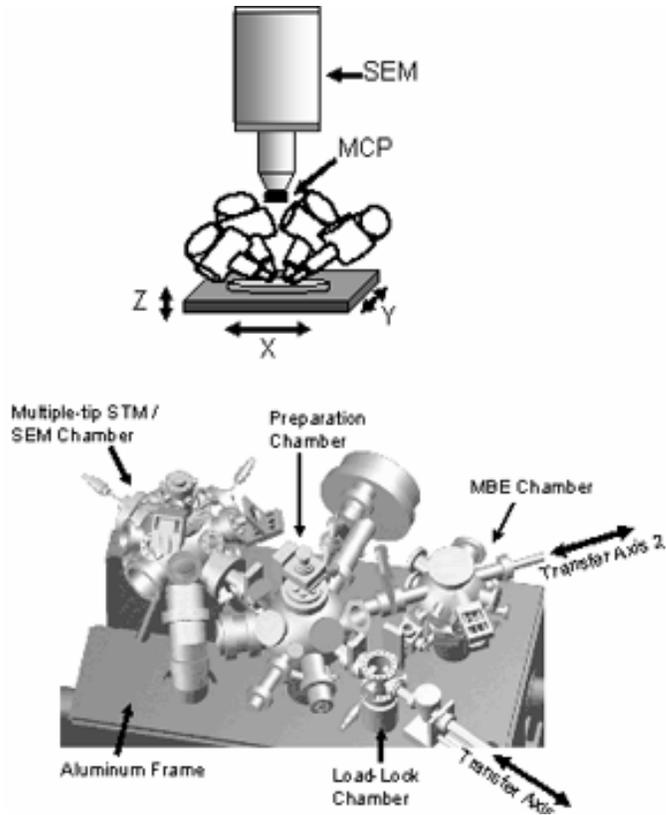

Fig. 1
O. Guise, et al.

# Surface Analysis and Preparation Chamber

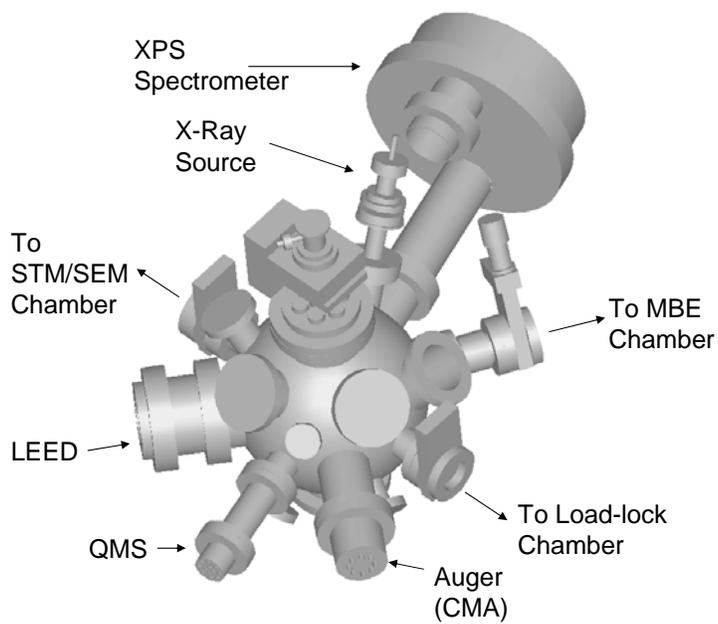

Schematic View of UHV Apparatus for Studies of Semiconductor Surfaces

- XPS Spectrometer
- X-Ray Source
- To STM/SEM Chamber
- To MBE Chamber
- LEED
- To Load-lock Chamber
- QMS
- Auger (CMA)

Fig. 2
O. Guise, et al.

# Molecular Beam Epitaxy Chamber

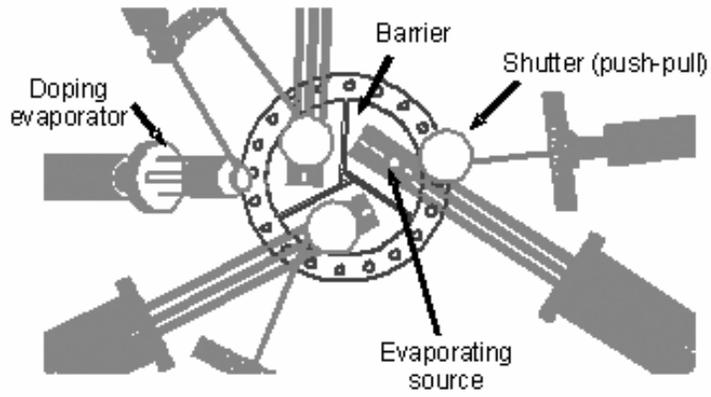

Fig. 3
O. Guise, et al.

# Sample Box System

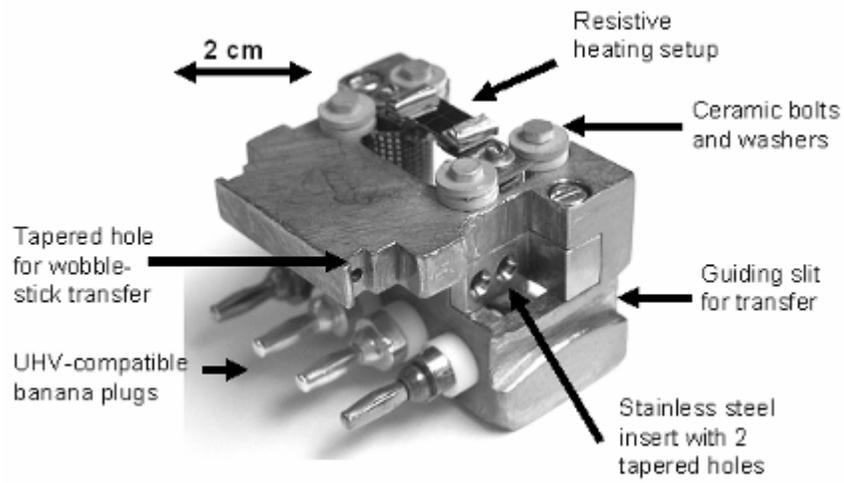

Fig. 4
O. Guise, et al.

# Sample Box Transfer Sequence

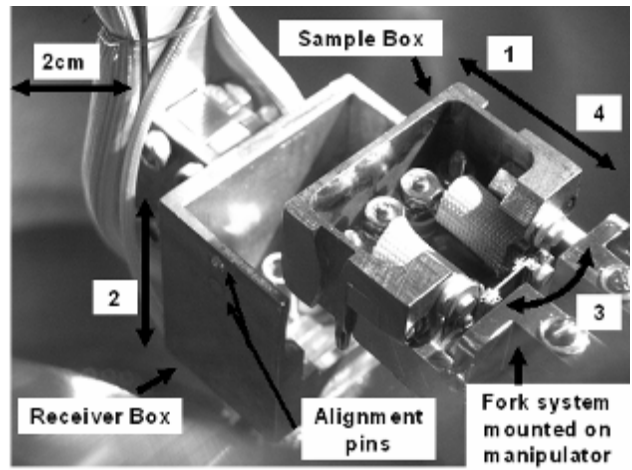

Fig. 5
O. Guise, et al.

# Multiple Tip STM Stage

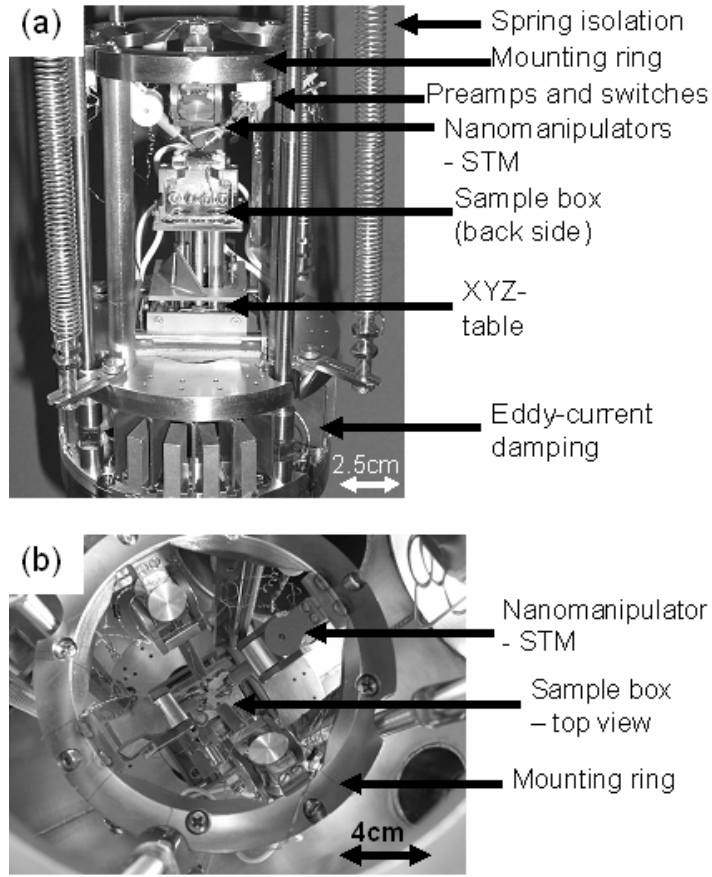

Fig. 6
O. Guise, et al.

# MM3a Nanomanipulator

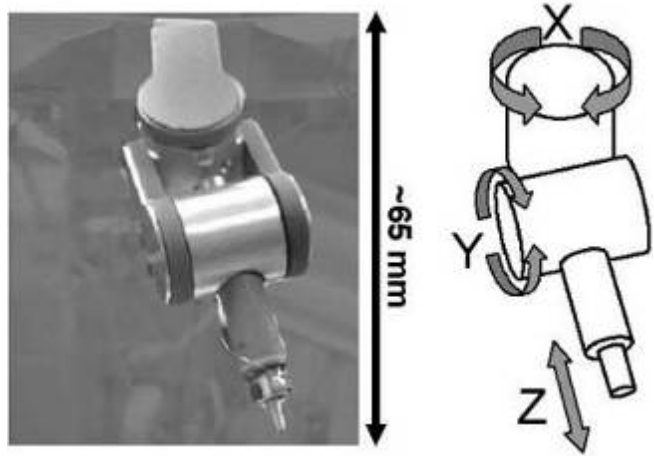

Fig. 7
O. Guise, et al.

# Preamplifier and Relay Network

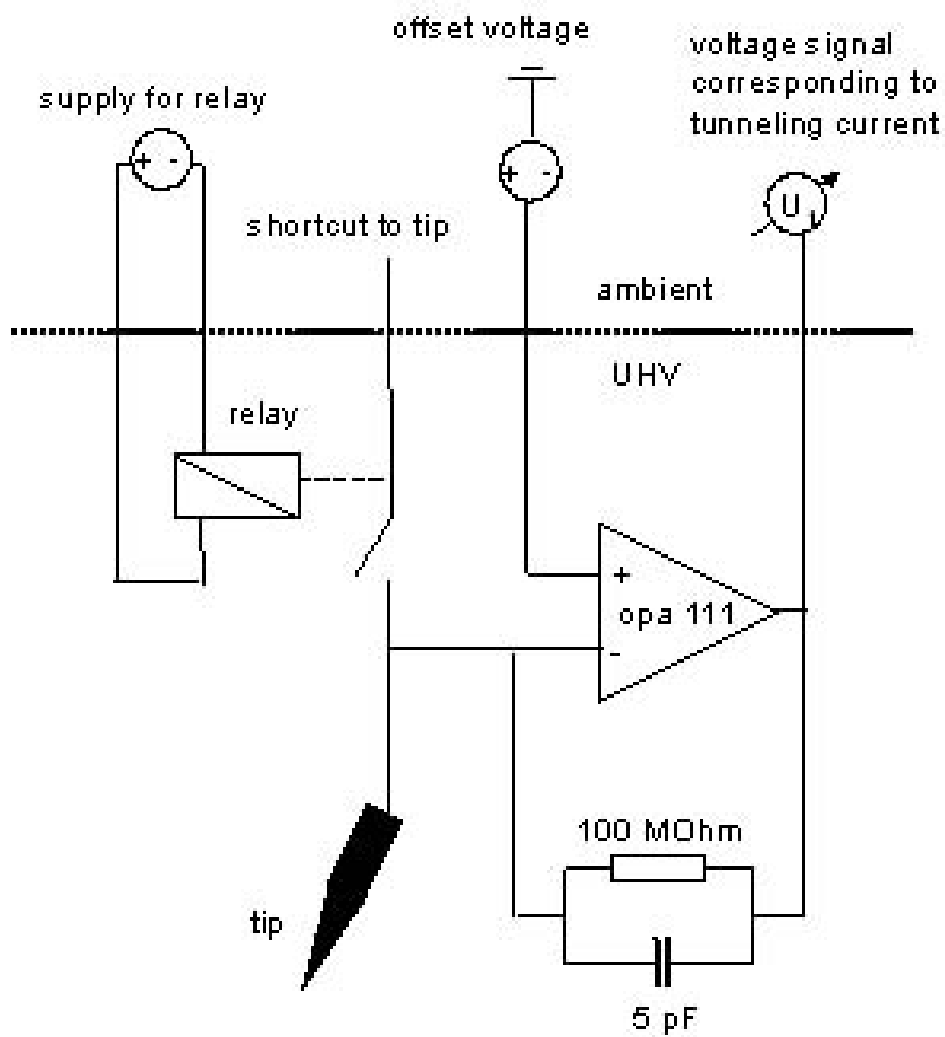

Fig. 8
O. Guise, et al.

# Schematics of the Tip Holder Assembly

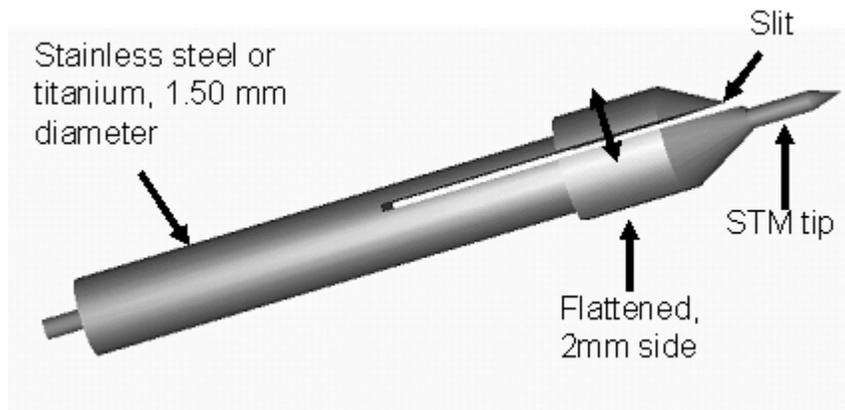

Fig. 9
O. Guise, et al.

# SEM – MCP Detection System

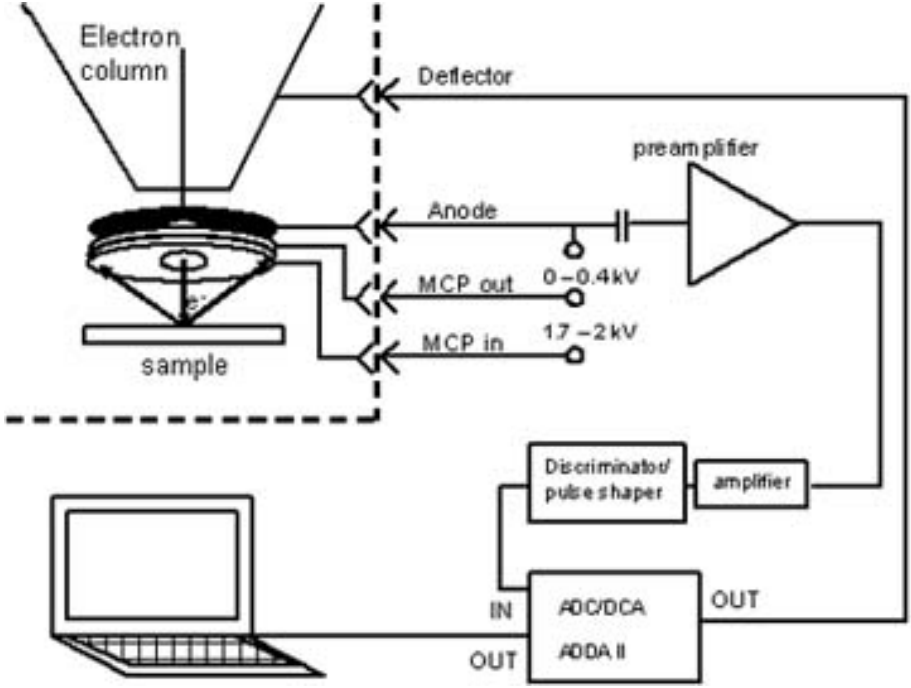

Fig. 10
O. Guise, et al.

# Schematic of the docking stage in the multiple-tip STM / SEM chamber

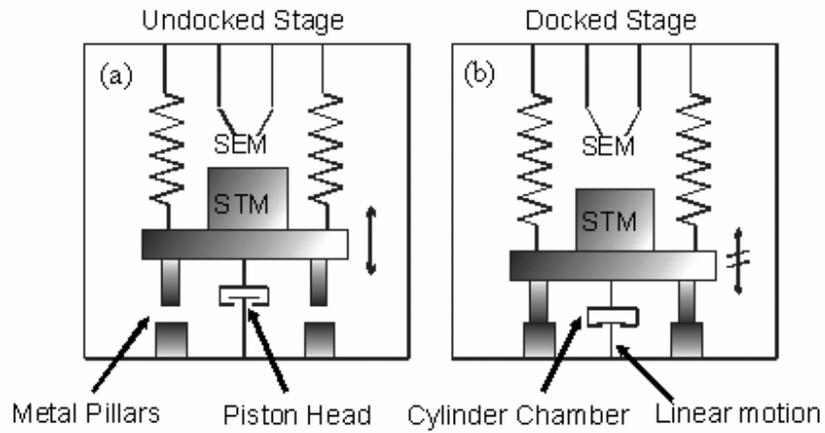

Fig. 11
O. Guise, et al.

# Vibration Analysis – Effect of the Eddy-current Damping System

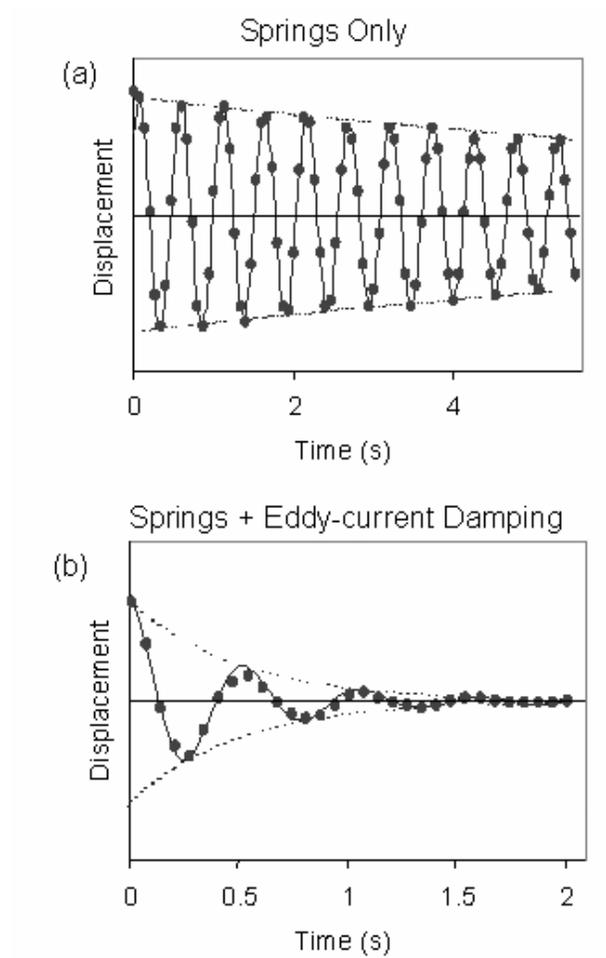

Fig. 12
O. Guise, et al.

# Reproducible STM Imaging of an Atomic Step on Graphite with Nanomanipulator

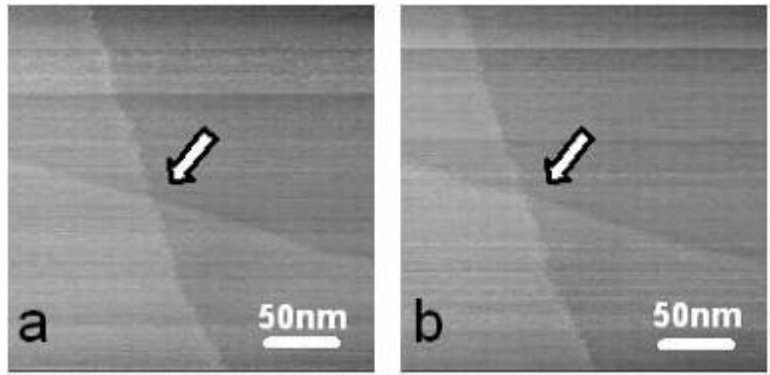

Fig. 13
O. Guise, et al.

# Four-point-probe Configuration

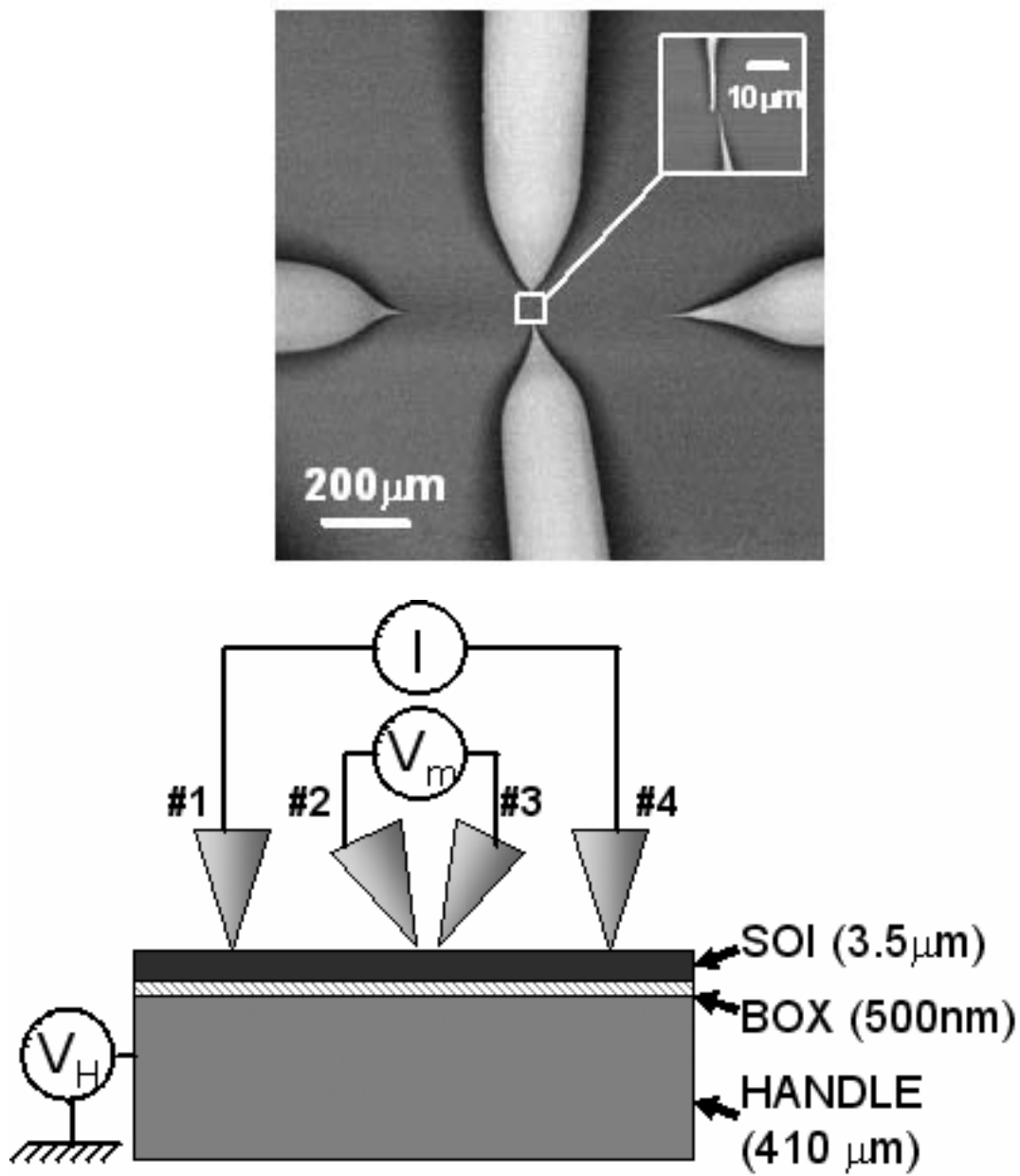

Fig. 14
O. Guise, et al.

# Field Effect Transistor Effect on SOI Crystal

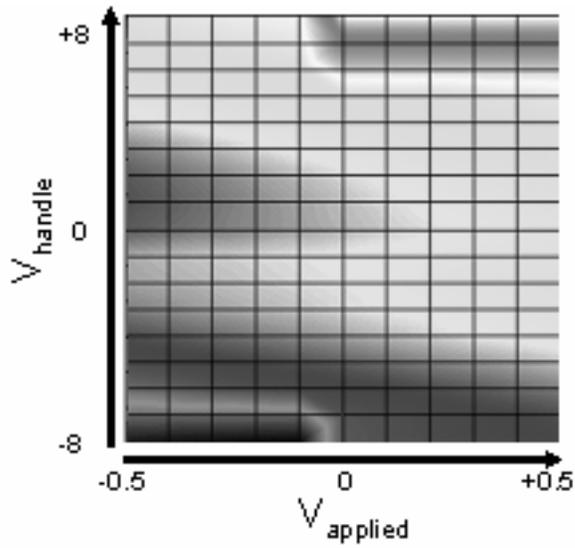

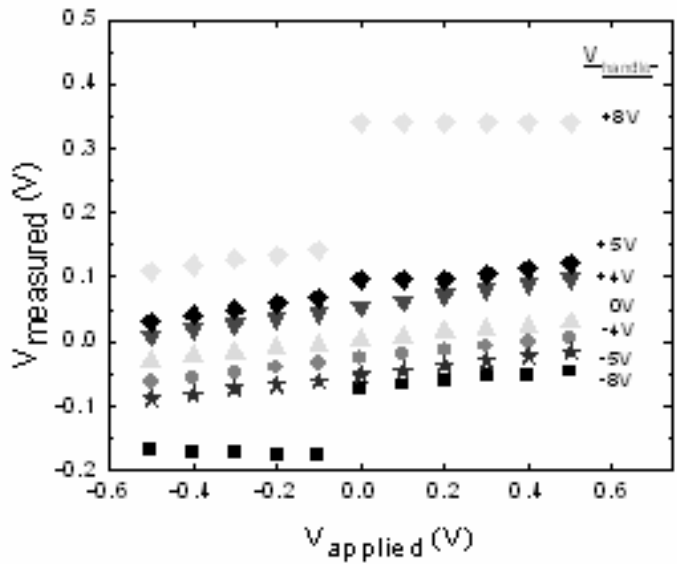

Fig. 15
O. Guise, et al.